\documentclass[11pt, fleqn]{article}
\usepackage{preamble} 
\usepackage{macros}
\usepackage{letterfonts}
\usepackage{commands}

\newcommand{\Glory}{\textsc{Glory}}
\newcommand{\Gnash}{\textsc{Gnash}}
\newcommand{\hubweight}{\mathrm{hub\_weight}}
\newcommand{\restweight}{\mathrm{rest\_weight}}
\newcommand{\maxrest}{\mathrm{max\_rest}}

\newcommand{\scoreG}{\mathrm{score}_{\Glory}}
\newcommand{\scoreN}{\mathrm{score}_{\Gnash}}
\newcommand{\nextheaven}{\mathrm{next}_{\mathrm{heaven}}}
\title{Heaven \& Hell: One-Step Hub Consensus\thanks{GUDraBIT Research Technical Report TR-2025-01.}}
\author{
  Nnamdi Aghanya \\ \texttt{nnamdi.aghanya@cranfield.ac.uk}
}
\date{}
\begin{document}
\maketitle
\begin{abstract}
\noindent In networked systems, enforcing a global configuration from a central authority against local peer influence is a foundational challenge. We investigate a model of influence dynamics on a finite weighted digraph where a distinguished hub node seeks to impose a desirable state, \Glory, upon the entire network. Our central objective is to identify a sharp, local, and efficiently verifiable graph-theoretic condition that guarantees global convergence to \Glory\ in a single synchronous update, irrespective of the initial network state. Our primary contribution is an exact necessary-and-sufficient condition based on the relative weights of incoming edges from the hub versus all other nodes. We specialise this to derive a sharp threshold for systems with a uniform-hub broadcast weight. The analysis is extended to a biased update rule and to asynchronous (Gauss–Seidel) updates, where we show one pass suffices under the same domination hypothesis. Every theoretical result is formally verified in the Coq proof assistant.
\end{abstract}

\section{Introduction}
\label{sec:intro}
In networked systems, achieving global consensus or enforcing a desired configuration is a foundational challenge, particularly when nodes are subject to conflicting local influences. We consider a scenario where a designated control node, or \texttt{hub}, attempts to steer the entire system towards a specific state (termed \Glory) against the potentially opposing influence of other nodes, which may promote an alternative state, \Gnash. This model's diverse phenomena range from a trusted server disseminating correct software versions in a peer-to-peer network to a central bank managing inflation expectations among regional actors.

Our central objective is to identify a sharp, local, and efficiently verifiable graph-theoretic condition that guarantees global convergence to \Glory\ in a single synchronous update, irrespective of the initial network state. Such a one-step guarantee is crucial for applications demanding rapid and reliable configuration control.

Classical binary-state dynamics include voter/majority models~\cite{Vazquez2008,Lambiotte2007}, linear-threshold processes~\cite{Kempe2003}, and Glauber-type updates on Ising graphs~\cite{Glauber1963}. Those lines typically study asymptotic behaviour, spectral or energy-based criteria, or average-case initialisations. Control-theoretic work on distributed consensus (and Byzantine-robust variants) focuses on linear averaging or fault bounds~\cite{Ren2008,Lamport1982}, again with multi-step convergence. Our setting differs in three respects: a non-linear, score-based update; an adversarial (worst-case) initial state; and an exact one-step criterion. The condition is vertex-local (hub-vs-rest inequality on in-weights), needs no spectral computation, and applies to arbitrary non-negative weights.

\begin{figure}[H]
\centering
\begin{adjustbox}{width=0.45\linewidth}
\begin{tikzpicture}[
    >=stealth,
    vertex/.style={circle, draw, very thick, minimum size=16pt, inner sep=0pt, font=\small},
    edgeu/.style={line width=0.9pt, draw=gray!60},
    every label/.style={font=\scriptsize, inner sep=1.5pt}
]
\definecolor{glory}{HTML}{2563EB}
\definecolor{gnash}{HTML}{EF4444}
\definecolor{hub}{HTML}{D97706}

\begin{scope}[shift={(0,0)}]
\node at (0,3.3) {\textbf{(a) Initial state $s_0$}};
\node[vertex, fill=glory!18, draw=glory, label=above:$v_1$] (a1) at (0,1.2) {};
\node[vertex, fill=gnash!18, draw=gnash, label=right:$v_2$] (a2) at (1.2,0) {};
\node[vertex, fill=glory!18, draw=glory, label=below:$v_3$] (a3) at (0,-1.2) {};
\node[vertex, fill=gnash!18, draw=gnash, label=left:$v_4$] (a4) at (-1.2,0) {};
\draw[edgeu] (a1) -- (a2) -- (a3) -- (a4) -- (a1);
\end{scope}

\begin{scope}[shift={(5.2,0)}]
\node at (0,3.3) {\textbf{(b) Next state $s_1$}};
\node[vertex, fill=gnash!18, draw=gnash, label=above:$v_1$] (b1) at (0,1.2) {};
\node[vertex, fill=glory!18, draw=glory, label=right:$v_2$] (b2) at (1.2,0) {};
\node[vertex, fill=gnash!18, draw=gnash, label=below:$v_3$] (b3) at (0,-1.2) {};
\node[vertex, fill=glory!18, draw=glory, label=left:$v_4$] (b4) at (-1.2,0) {};
\draw[edgeu] (b1) -- (b2) -- (b3) -- (b4) -- (b1);
\end{scope}

\draw[->, very thick] (1.2,-1.8) to[bend right=15] node[below, pos=0.5, font=\scriptsize]{synchronous update} (4.0,-1.8);
\draw[->, very thick] (4.0,1.8) to[bend right=15] node[above, pos=0.5, font=\scriptsize]{synchronous update} (1.2,1.8);
\end{tikzpicture}
\end{adjustbox}
\caption{\texttt{HellC4}: A synchronous 2-cycle. The two parity configurations, (a) and (b), alternate with each update step.}
\label{fig:hellc4-twocycle}
\end{figure}
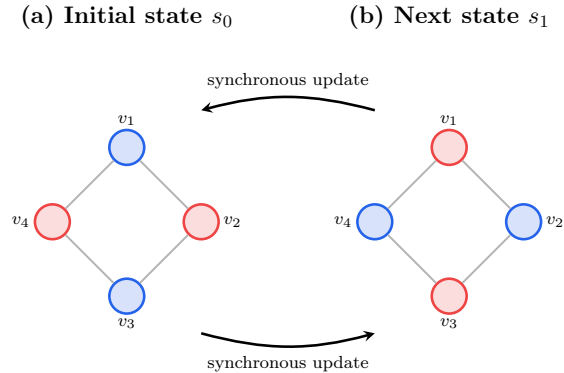

This perspective is also tied to influence and seeding problems~\cite{Kempe2003}. Linear-threshold cascades and influence maximisation address reachability over time under budgeted interventions; here, the \texttt{budget} is concentrated at a single hub, whether one step suffices for universal adoption. The failure mode without a hub is immediate (two cycles under parity on a 4-cycle; see \autoref{fig:hellc4-twocycle}); the hub breaks such cycles once its inbound weight into each non-hub exceeds the competing inbound from the rest. The empirical sweep in \autoref{fig:empirical_plot} matches the predicted threshold $\maxrest$.

Our contributions are as follows:
\begin{enumerate}[label=(\roman*), nolistsep]
\item We establish an exact necessary-and-sufficient condition for one-step convergence, expressed as a simple inequality between hub and non-hub influence at each vertex (\autoref{thm:main_equiv}).
\item This general result is specialised to the practical case of a uniform-hub broadcast, yielding a sharp threshold characterised by a single global graph parameter, $\maxrest$ (\autoref{thm:uniform_threshold}).
\item We extend our analysis to include a biased ($\tau$) update rule, providing a correspondingly adjusted convergence condition (\autoref{thm:tau_equiv}).
\item We demonstrate that the same hub domination hypothesis guarantees convergence in one pass under asynchronous (Gauss--Seidel) updates (\autoref{thm:async_one_pass}).
    \item Finally, every theoretical result presented is backed by a formal Coq development, aligning each theorem with a mechanised proof and ensuring its correctness.~\footnote{Coq Files can be found \href{https://github.com/gudfit/HeavenHell}{here}.}
\end{enumerate}

\newpage
\section{Methodology}
\label{sec:Methodology}

\paragraph{Model and Update Rules.}
We consider a finite vertex set $\mathcal{V}$ with a distinguished hub $g \in \mathcal{V}$. The network is a weighted directed graph where edge weights are given by a function $w: \mathcal{V} \times \mathcal{V} \to \mathbb{N}$. Each vertex $v \in \mathcal{V}$ can be in one of two states from the set $\{\Glory, \Gnash\}$. The system's configuration is a state function $s: \mathcal{V} \to \{\Glory, \Gnash\}$.

For any vertex $v$, its incoming influence from peers in a state $x$ is quantified by a score function, $\mathrm{score}_x(s,v) \coloneqq \sum_{u \in \mathcal{V}} w(u,v) \cdot \mathbb{I}(s(u) = x)$, where $\mathbb{I}(\cdot)$ is the indicator function.

The update rule prioritises the hub's state. We define a hub-forced state $s_g$, where $s_g(v) = \Glory$ if $v=g$ and $s_g(v)=s(v)$ otherwise. A vertex $v$ updates its state based on a majority vote of incoming influence in this hub-forced state. The unbiased update rule, $\nextheaven$, is defined as:
\[
\nextheaven(s, v) \coloneqq
\begin{cases}
    \Glory & \text{if } v = g \\
    \Glory & \text{if } v \neq g \text{ and } \scoreG(s_g, v) \ge \scoreN(s_g, v) \\
    \Gnash & \text{if } v \neq g \text{ and } \scoreG(s_g, v) < \scoreN(s_g, v)
\end{cases}
\]
The $\tau$-biased variant, $\nextheaven^{\tau}$, introduces a non-negative bias $\tau(v)$ to the \Glory\ score, thereby comparing $\scoreG(s_g, v) + \tau(v)$ against $\scoreN(s_g, v)$.

For analysis, we partition the total incoming weight at any vertex $v$ into that from the hub, $\hubweight(v) \coloneqq w(g,v)$, and that from all other nodes, $\restweight(v) \coloneqq \sum_{u \neq g} w(u,v)$. The foundational identities and bounds relating these weights to the score functions are derived in \autoref{sec:appendix_aux}.

\subsection{Main Theorems}
\label{sec:main}

We present our primary results concerning the conditions for guaranteed one-step convergence. The proofs are based on the lemmas and score properties established in \autoref{sec:appendix_aux}.

\begin{theorem}[One-step Convergence, Unbiased]
\label{thm:main_equiv}
Global convergence to \Glory\ in one synchronous step, from any initial state, is guaranteed if and only if the hub's influence dominates the collective non-hub influence at every other vertex:
\[ (\forall s, v, \ \nextheaven(s, v) = \Glory) \iff (\forall v \neq g, \ \hubweight(v) \ge \restweight(v)) \]
\end{theorem}
\begin{proof}
The proof is deferred to the appendix, see \autoref{thm:global_conv_equiv}.
\end{proof}

\begin{theorem}[Uniform Hub Threshold]
\label{thm:uniform_threshold}
If the hub influence is uniform, i.e., $\hubweight(v)=W$ for all $v \neq g$, and we define $\maxrest \coloneqq \max_{v \neq g} \restweight(v)$, then one-step convergence holds if and only if $W \ge \maxrest$.
\end{theorem}
\begin{proof}
By \autoref{thm:main_equiv}, the system converges if and only if $(\forall v \neq g, \hubweight(v) \ge \restweight(v))$. Substituting the uniform hub weight, this becomes $(\forall v \neq g, W \ge \restweight(v))$. By \autoref{lem:max_rest_equiv}, this universal condition is equivalent to $W$ being greater than or equal to the maximum of all such rest weights, which is precisely $W \ge \maxrest$.
\end{proof}

\begin{theorem}[Sharpness of the Threshold]
\label{thm:sharpness}
If the uniform hub weight $W$ is below the threshold, $W < \maxrest$, there exists a non-hub vertex $v$ for which the update from the \texttt{all\_gnash} state results in \Gnash.
\end{theorem}
\begin{proof}
Assume $W < \maxrest$. By the property of the maximum function (\autoref{lem:max_rest_gt_exists}), there must exist a vertex $v_0 \neq g$ such that $\restweight(v_0) > W$. Given the uniform hub weight, this implies $\hubweight(v_0) < \restweight(v_0)$. By \autoref{lem:instability_weak_hub}, this is the precise condition under which $\nextheaven(\texttt{all\_gnash}, v_0) = \Gnash$.
\end{proof}

\begin{theorem}[One-step Convergence, $\tau$-Biased]
\label{thm:tau_equiv}
For the biased update rule, one-step convergence holds if and only if the sum of the hub's influence and the local bias dominates the rest-influence:
\[ (\forall s, v, \ \nextheaven^{\tau}(s, v) = \Glory) \iff (\forall v \neq g, \ \hubweight(v) + \tau(v) \ge \restweight(v)) \]
\end{theorem}
\begin{proof}
The proof follows the same structure as that of \autoref{thm:main_equiv} and is presented in the appendix as part of \autoref{thm:global_conv_equiv}.
\end{proof}

\section{Algorithmic Certification and Threshold Extraction}
\label{sec:algorithm}

The theoretical conditions translate into efficient algorithms for certifying a network's stability or extracting the minimum required hub strength.

\paragraph{Linear-time certification.}
The domination condition can be checked in $O(|E|)$ time and $O(|\mathcal V|)$ space. A single pass over the edges is sufficient: initialise arrays for hub and rest weights, iterate through edges to populate them, and finally verify the inequality for all non-hub vertices. This procedure is formalised in \autoref{alg:certify}.

\begin{algorithm}[H]
\caption{Certify one-step consensus and compute thresholds}
\label{alg:certify}
Initialise arrays $\text{hub\_weight}[\cdot]\gets 0$, $\text{rest\_weight}[\cdot]\gets 0$ on $\mathcal V$\;
\ForEach{edge $(u,v)\in E$ with weight $w$}{
  \If{$u=g$}{
    $\text{hub\_weight}[v] \gets \text{hub\_weight}[v]+w$\;
  }
  \Else{
    $\text{rest\_weight}[v] \gets \text{rest\_weight}[v]+w$\;
  }
}
\KwRet{\textsc{Pass} iff $\text{hub\_weight}[v]\ge \text{rest\_weight}[v]$ for all $v\neq g$}
\end{algorithm}

\paragraph{Uniform-hub threshold.}
If the hub broadcasts a uniform weight $W$ to all $v\neq g$, the minimal one-step convergence weight $W^\*$ is $W^\* = \max_{v\neq g} \restweight(v)$. This value is a natural by-product of the certification process in \autoref{alg:certify}.

\paragraph{Tie-bias \texorpdfstring{$\tau$}{tau}.}
For the $\tau$-biased rule, the necessary and sufficient condition is $\forall v\neq g: \hubweight(v)+\tau(v) \ge \restweight(v)$. With a uniform hub weight $W$, the minimal broadcast weight is $W^\*_\tau = \max_{v\neq g} (\restweight(v)-\tau(v))_+$, where $(\cdot)_+$ denotes truncation at zero. This requires a single pass to compute rest weights, followed by a pass to find the maximum of the needed values.

\paragraph{Seeding in two steps.}
If a seed set $S\subseteq \mathcal V\setminus\{g\}$ of vertices is preset to \Glory, a sufficient condition for these vertices to remain in \Glory\ is $\hubweight(v) + w_{\text{from}}(S,v) \ge \text{rest\_outside}(S,v)$ for all $v\in S$, where $w_{\text{from}}(S,v)=\sum_{u\in S} w(u,v)$ and $\text{rest\_outside}(S,v)=\sum_{u\notin S\cup\{g\}} w(u,v)$. Both sums are computable in $O(|E|)$ time.

\paragraph{Asynchronous schedule.}
Under the standard domination condition, any asynchronous update schedule that includes each non-hub vertex at least once guarantees convergence to \Glory\ after a single pass (\autoref{thm:async_one_pass}).

\paragraph{Global coarse bound.}
A quick a priori upper bound on the required uniform weight $W^\*$ can be derived without a full weight scan. Let $\deg^{\text{in}}(v)$ be the number of non-hub in-neighbours of $v$ and $w_{\max}$ be the maximum non-hub in-edge weight. Then $\restweight(v) \le \deg^{\text{in}}(v)\cdot w_{\max}$, which implies $W^\* \le (\max_{v\neq g} \deg^{\text{in}}(v))\cdot w_{\max}$.

\section{Examples}
\label{sec:examples}
We illustrate the dynamics with two minimal examples. Detailed computations and formal proofs are in \autoref{sec:appendix_examples}.

\subsection{\texttt{HellC4}}
\label{subsec:hellc4_main}
A 4-cycle with unit-weight edges and no hub exhibits oscillatory behaviour. From a checkerboard initial state, the system enters a two-cycle, flipping between the two possible parity configurations in each synchronous step. This demonstrates the failure to achieve consensus without a sufficiently strong coordinating influence. A visualisation of this two-cycle is presented in \autoref{fig:hellc4-twocycle}.

\subsection{\texttt{HeavenExample}}
\label{subsec:heavenexample_main}
\begin{figure}[H]
\centering
\begin{adjustbox}{max width=0.65\linewidth}
\begin{tikzpicture}[
    >=stealth,
    vertex/.style={circle, draw, very thick, minimum size=18pt, inner sep=0pt, font=\small},
    edge1/.style={line width=0.9pt, draw=gray!60},
    edgeW/.style={-{Stealth[length=2.2mm]}, line width=1.1pt, draw=hub!90},
    every label/.style={font=\scriptsize, inner sep=1.5pt}
]
\definecolor{hub}{HTML}{D97706}

\node[vertex, label=above:$v_1$] (v1) at (0,1.8) {};
\node[vertex, label=right:$v_2$] (v2) at (1.8,0) {};
\node[vertex, label=below:$v_3$] (v3) at (0,-1.8) {};
\node[vertex, label=left:$v_4$] (v4) at (-1.8,0) {};
\draw[edge1] (v1) -- (v2) -- (v3) -- (v4) -- (v1);

\node[vertex, fill=hub!12, draw=hub!90, label={[font=\scriptsize]right:{$h=g$}}] (h) at (0,0) {\scriptsize hub};

\draw[edgeW] (h) -- node[pos=0.55, above, sloped, font=\tiny] {$W$} (v1);
\draw[edgeW] (h) -- node[pos=0.55, right, sloped, font=\tiny] {$W$} (v2);
\draw[edgeW] (h) -- node[pos=0.45, below, sloped, font=\tiny] {$W$} (v3);
\draw[edgeW] (h) -- node[pos=0.45, left, sloped, font=\tiny] {$W$} (v4);

\node[font=\scriptsize, align=left, anchor=west] at (2.6,0.6) {$\text{hub\_weight}(v_i)=W$};
\node[font=\scriptsize, align=left, anchor=west] at (2.6,-0.1) {$\text{rest\_weight}(v_i)=1+1=2$};
\end{tikzpicture}
\end{adjustbox}
\caption{\texttt{HeavenExample}: A 4-cycle with a hub $h=g$ directing influence of weight $W$ to each non-hub vertex. The cycle edges have unit weight, so $\text{rest\_weight}(v_i)=2$.}
\label{fig:heaven-graph}
\end{figure}
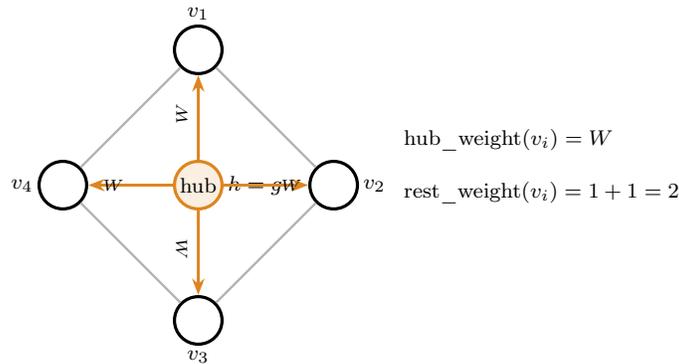

This example augments the 4-cycle with a central hub $g$ that has outgoing edges of uniform weight $W$ to each other vertex (see \autoref{fig:heaven-graph}). The rest-weight for each non-hub vertex is 2. Applying our main theorems yields a sharp threshold:
\begin{itemize}[nolistsep]
    \item If $W < 2$, the condition of \autoref{thm:uniform_threshold} is not met. \autoref{thm:sharpness} predicts a failure to converge from \texttt{all\_gnash}, which is confirmed by direct computation (\autoref{thm:heaven_counterexample}).
    \item If $W \ge 2$, the condition is met, and \autoref{thm:uniform_threshold} guarantees one-step convergence to \Glory\ from any initial state (\autoref{thm:heaven_convergence}).
\end{itemize}
This simple case directly illustrates the sharpness of our derived threshold.

\section{Empirical Validation}
\label{sec:experiments}
To validate the sharpness of the uniform hub threshold predicted by \autoref{thm:uniform_threshold}, we conducted numerical simulations. The procedure involved generating random directed graphs of 50 vertices with a designated hub. Non-hub vertices were connected with an edge probability of 0.1, and edge weights were drawn uniformly from $\{1, \dots, 10\}$.

For each graph, we computed its theoretical threshold, $\maxrest$. We then swept the uniform hub weight, $W$, across a range of values. For each $W$, we simulated one synchronous update step from the \texttt{all\_gnash} configuration and measured the fraction of non-hub vertices converging to \Glory. We also evaluated the success rate of one-pass asynchronous updates over 100 trials with randomised schedules.

\begin{figure}[H]
\centering
\includegraphics[width=0.82\linewidth]{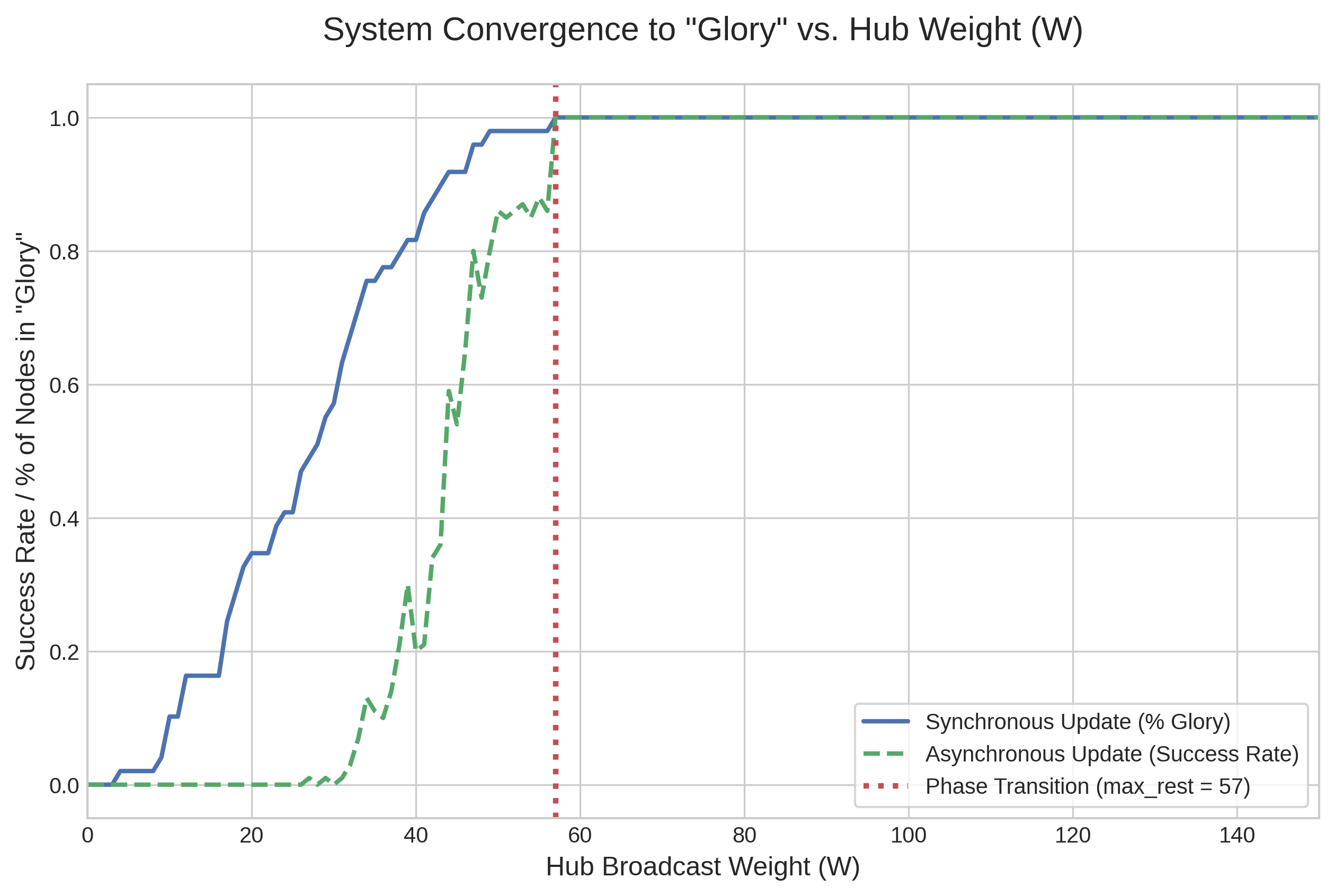}
\caption{Empirical sweep of uniform hub weight $W$ versus one-step convergence rate for a randomly generated 50-vertex graph. The observed phase transition occurs precisely at the calculated theoretical threshold of $W = \maxrest = 57$.}
\label{fig:empirical_plot}
\end{figure}

The results, exemplified in \autoref{fig:empirical_plot}, confirm the theoretical predictions. The synchronous update exhibits a stark phase transition: the convergence rate is partial or zero for all $W < \maxrest$ and jumps to 100\% precisely at $W = \maxrest$. For the instance shown, the calculated threshold was $\maxrest = 57$, which is exactly where the transition occurs. The asynchronous updates also achieve a 100\% success rate for all $W \ge \maxrest$, validating the one-pass guarantee of \autoref{thm:async_one_pass}. This evidence supports the sharpness and correctness of the derived domination conditions.

\section{Conclusion}
\label{sec:conclusion}
We have established a sharp, necessary and sufficient graph-theoretic condition for a central hub to enforce a global state of \Glory\ in a single synchronous update. This condition, $\hubweight(v) \ge \restweight(v)$ for all non-hubs $v$, is intuitive, analytically powerful, and algorithmically efficient to verify. We characterised its implications for uniform-hub systems, biased updates, and asynchronous dynamics, demonstrating the robustness of the core domination principle. All claims are supported by mechanically-checked proofs in Coq, providing the highest standard of formal assurance. Future work could explore multi-step convergence guarantees when the one-step condition is not met, or investigate the impact of stochastic noise on the update rule.

\newpage
\bibliography{ref}
\pagebreak
\appendix

\section{Auxiliary Proofs and Derivations}\label{sec:appendix_aux}

\subsection{Score and Weight Structure}
This section establishes the core relationships between total weight, hub weight, rest weight, and the resulting scores under the hub-forced state $s_g$.

\begin{lemma}[Total Weight Decomposition]
\label{lem:total_weight_split}
For any vertex $v$, the total incoming weight decomposes as:
\[ \sum_{u \in \mathcal{V}} w(u,v) = \text{hub\_weight}(v) + \text{rest\_weight}(v) \]
\end{lemma}
\begin{proof}
The sum over all vertices $\mathcal{V}$ can be partitioned into the term for $u=g$ and the sum over all $u \neq g$. The former is $w(g,v) \equiv \text{hub\_weight}(v)$ and the latter is $\sum_{u \neq g} w(u,v) \equiv \text{rest\_weight}(v)$ by definition.
\end{proof}

\begin{lemma}[Glory Score Decomposition]
\label{lem:scoreG_split}
For any state $s$ and vertex $v$, the \textsc{Glory} score under $s_g$ decomposes as:
\[ \text{score}_{\textsc{Glory}}(s_g, v) = \text{hub\_weight}(v) + \sum_{u \neq g} w(u,v) \cdot \mathbb{I}(s(u) = \textsc{Glory}) \]
\end{lemma}
\begin{proof}
The sum for $\text{score}_{\textsc{Glory}}(s_g, v)$ is split into the term for $u=g$ and the sum over $u \neq g$. For $u=g$, the term is $w(g,v) \cdot \mathbb{I}(s_g(g)=\textsc{Glory}) = w(g,v) = \text{hub\_weight}(v)$. For all $u \neq g$, $s_g(u) = s(u)$, which yields the second term.
\end{proof}

\begin{lemma}[Fundamental Score Bounds]
\label{lem:bounds_force_g}
For any state $s$ and vertex $v$:
\begin{enumerate}[nolistsep, label=(\roman*)]
    \item $\text{score}_{\textsc{Gnash}}(s_g, v) \le \text{rest\_weight}(v)$.
    \item $\text{hub\_weight}(v) \le \text{score}_{\textsc{Glory}}(s_g, v)$.
\end{enumerate}
\end{lemma}
\begin{proof}\hfill
\begin{enumerate}[nolistsep, label=(\roman*)]
    \item The sum for $\text{score}_{\textsc{Gnash}}(s_g, v)$ is $\sum_{u \in \mathcal{V}} w(u,v) \cdot \mathbb{I}(s_g(u) = \textsc{Gnash})$. The term for $u=g$ is zero as $s_g(g)=\textsc{Glory}$. For $u \neq g$, the indicator is at most 1, so the sum is bounded by $\sum_{u \neq g} w(u,v) = \text{rest\_weight}(v)$.
    \item The sum for $\text{score}_{\textsc{Glory}}(s_g, v)$ includes the term for $u=g$, which is $w(g,v) \cdot \mathbb{I}(\textsc{Glory}=\textsc{Glory}) = \text{hub\_weight}(v)$. All other terms are non-negative.
\end{enumerate}
\end{proof}

\begin{lemma}[Scores from the \texttt{all\_gnash} State]
\label{lem:scores_all_gnash}
For any vertex $v$, let $s_{ag} = \text{force\_g}(\text{all\_gnash})$. Then:
\begin{enumerate}[nolistsep, label=(\roman*)]
    \item $\text{score}_{\textsc{Glory}}(s_{ag}, v) = \text{hub\_weight}(v)$.
    \item $\text{score}_{\textsc{Gnash}}(s_{ag}, v) = \text{rest\_weight}(v)$.
\end{enumerate}
\end{lemma}
\begin{proof}\hfill
\begin{enumerate}[nolistsep, label=(\roman*)]
    \item Using \autoref{lem:scoreG_split}, the non-hub term is $\sum_{u \neq g} w(u,v) \cdot \mathbb{I}(\text{all\_gnash}(u) = \textsc{Glory})$. As $\text{all\_gnash}(u)=\textsc{Gnash}$, this sum is zero.
    \item The sum for $\text{score}_{\textsc{Gnash}}(s_{ag}, v)$ has a zero term for $u=g$. For all $u \neq g$, $s_{ag}(u) = \text{all\_gnash}(u) = \textsc{Gnash}$, so the indicator is 1. The sum becomes $\sum_{u \neq g} w(u,v) = \text{rest\_weight}(v)$.
\end{enumerate}
\end{proof}

\subsection{Update Rule Characterisations}

\begin{lemma}[Fixed State of the Hub]
For any state $s$, $\text{next}_{\text{heaven}}(s, g) = \textsc{Glory}$ and $\text{next}_{\text{heaven}}^{\tau}(s, g) = \textsc{Glory}$.
\end{lemma}
\begin{proof}
This is by definition of the update rules.
\end{proof}

\begin{lemma}[Update Rule Equivalence for Non-Hubs]
For any state $s$ and vertex $v \neq g$:
\begin{enumerate}[nolistsep, label=(\roman*)]
    \item $\text{next}_{\text{heaven}}(s, v) = \textsc{Glory} \iff \text{score}_{\textsc{Gnash}}(s_g, v) \le \text{score}_{\textsc{Glory}}(s_g, v)$.
    \item $\text{next}_{\text{heaven}}^{\tau}(s, v) = \textsc{Glory} \iff \text{score}_{\textsc{Gnash}}(s_g, v) \le \text{score}_{\textsc{Glory}}(s_g, v) + \tau(v)$.
\end{enumerate}
\end{lemma}
\begin{proof}
A direct translation of the update rule's logic. A vertex becomes \textsc{Gnash} only if the strict inequality holds against \textsc{Glory}'s score (plus bias). This lemma states the condition for the complementary case.
\end{proof}

\begin{theorem}[Domination Sufficiency]
\label{thm:domination_sufficiency}
For any state $s$ and vertex $v \neq g$, if $\text{rest\_weight}(v) \le \text{hub\_weight}(v)$, then $\text{next}_{\text{heaven}}(s, v) = \textsc{Glory}$. The same holds for the $\tau$-biased rule if $\text{rest\_weight}(v) \le \text{hub\_weight}(v) + \tau(v)$.
\end{theorem}
\begin{proof}
Using the bounds from \autoref{lem:bounds_force_g} and the domination hypothesis, we form the chain:
\[ \text{score}_{\textsc{Gnash}}(s_g, v) \le \text{rest\_weight}(v) \le \text{hub\_weight}(v) \le \text{score}_{\textsc{Glory}}(s_g, v). \]
The result follows from the update rule equivalence. The proof for the $\tau$-biased case is identical, carrying the $+\tau(v)$ term through the inequalities.
\end{proof}

\begin{lemma}[Domination Condition from \texttt{all\_gnash}]
\label{lem:domination_all_gnash}
For $v \neq g$, $\text{next}_{\text{heaven}}(\text{all\_gnash}, v) = \textsc{Glory} \iff \text{hub\_weight}(v) \ge \text{rest\_weight}(v)$. For the $\tau$-biased rule, the equivalence is \\ $\text{next}_{\text{heaven}}^{\tau}(\text{all\_gnash}, v) = \textsc{Glory} \iff \text{hub\_weight}(v) + \tau(v) \ge \text{rest\_weight}(v)$.
\end{lemma}
\begin{proof}
This follows by substituting the exact scores for the \texttt{all\_gnash} case from \autoref{lem:scores_all_gnash} into the update rule characterisation.
\end{proof}

\begin{theorem}[Global Convergence Equivalence]
\label{thm:global_conv_equiv}
Global one-step convergence to \textsc{Glory} from any state occurs if and only if the hub domination condition holds for all non-hub vertices.
\[ (\forall s, v, \text{next}_{\text{heaven}}(s, v) = \textsc{Glory}) \iff (\forall v \neq g, \text{hub\_weight}(v) \ge \text{rest\_weight}(v)) \]
The same equivalence holds for the $\tau$-biased system with its corresponding domination condition.
\end{theorem}
\begin{proof}
($\Leftarrow$) If the domination condition holds for all non-hubs, then by \autoref{thm:domination_sufficiency}, every non-hub vertex updates to \textsc{Glory}. The hub updates to \textsc{Glory} by definition.
($\Rightarrow$) If convergence happens from any state, it must happen from the worst-case state \texttt{all\_gnash}. By \autoref{lem:domination_all_gnash}, this implies the domination condition must hold for every non-hub vertex.
\end{proof}

\subsection{Uniform-Hub Thresholds}

\begin{lemma}[Global vs Local Rest Weight Bounds]
\label{lem:max_rest_equiv}
For any threshold $W$, $\maxrest \le W \iff (\forall v \neq g, \text{rest\_weight}(v) \le W)$.
\end{lemma}
\begin{proof}
($\Rightarrow$) If $\maxrest \le W$, then for any $v \neq g$, we have $\text{rest\_weight}(v) \le \maxrest \le W$.
($\Leftarrow$) If $\text{rest\_weight}(v) \le W$ for all non-hubs $v$, then $W$ is an upper bound for all elements in the set whose maximum is $\maxrest$. The maximum itself must therefore be at most $W$.
\end{proof}

\begin{lemma}[Instability from a Weak Hub]
\label{lem:instability_weak_hub}
For any $v \neq g$, if $\text{hub\_weight}(v) < \text{rest\_weight}(v)$, then $\text{next}_{\text{heaven}}(\text{all\_gnash}, v) = \textsc{Gnash}$.
\end{lemma}
\begin{proof}
This is the contrapositive of the forward direction of \autoref{lem:domination_all_gnash}.
\end{proof}

\begin{lemma}[Existence of a Dominating Rest Weight]
\label{lem:max_rest_gt_exists}
If $W < \maxrest$, there exists a non-hub vertex $v$ such that $\text{rest\_weight}(v) > W$.
\end{lemma}
\begin{proof}
This is a property of the maximum function. If the maximum of a set of values is greater than $W$, at least one value in that set must be greater than $W$.
\end{proof}

\subsection{Asynchronous Schedule}

\begin{theorem}[One-Pass Asynchronous Convergence]
\label{thm:async_one_pass}
If the hub domination condition $(\forall v\neq g, \text{hub\_weight}(v) \ge \text{rest\_weight}(v))$ holds, then any asynchronous update schedule that includes each non-hub vertex at least once will result in all non-hub vertices being in state \textsc{Glory} after one pass.
\end{theorem}
\begin{proof}
Let \texttt{sched} be such a schedule. Consider any non-hub vertex $v$. Since $v \in \texttt{sched}$, at some point in the schedule, $v$ is updated. The state of $v$ becomes $\text{next}_{\text{heaven}}(s', v)$ for some intermediate state $s'$. By the domination condition and \autoref{thm:domination_sufficiency}, this update always results in \textsc{Glory}, because the condition is independent of the current state $s'$. If $v$ is updated multiple times, each update will result in \textsc{Glory}. Thus, after the pass is complete, the state of $v$ must be \textsc{Glory}. This holds for all non-hub vertices.
\end{proof}

\subsection{Fixed Points}
Finally, we confirm that the state of unanimous \textsc{Glory} is a fixed point of the dynamics.

\begin{theorem}[Fixed Point at Unanimous Glory]
The state \texttt{all\_glory} is a fixed point for both the unbiased and $\tau$-biased update rules.
\end{theorem}
\begin{proof}
For the state $s=\text{all\_glory}$, the hub-forced state is also \texttt{all\_glory}. For any vertex $v$, the score for \textsc{Gnash} is $\text{score}_{\textsc{Gnash}}(s_g,v)=0$, as no vertex is in state \textsc{Gnash}. The score for \textsc{Glory} is non-negative. Thus, the condition for updating to \textsc{Glory} is always met. The hub also remains \textsc{Glory} by definition.
\end{proof}

\newpage
\section{Example Computations}
\label{sec:appendix_examples}

\subsection{\texttt{HellC4}}
This model consists of a 4-cycle graph $\mathcal{V}_4 = \{v_1, v_2, v_3, v_4\}$ with unit weights between adjacent vertices and no hub. The update rule is a standard majority vote.

\begin{theorem}[Two-Cycle Oscillation]
For the state $s_0 = (\textsc{Glory}, \textsc{Gnash}, \textsc{Glory}, \textsc{Gnash})$, one synchronous update yields $s_1 = (\textsc{Gnash}, \textsc{Glory}, \textsc{Gnash}, \textsc{Glory})$, and a further update returns to $s_0$.
\end{theorem}
\begin{proof}
For state $s_0$:
\begin{itemize}[nolistsep]
    \item At $v_1$: Neighbours $v_2, v_4$ are \textsc{Gnash}. Score vector is $(\text{score}_{\textsc{Glory}}, \text{score}_{\textsc{Gnash}}) = (0, 2)$. Next state: \textsc{Gnash}.
    \item At $v_2$: Neighbours $v_1, v_3$ are \textsc{Glory}. Score vector is $(2, 0)$. Next state: \textsc{Glory}.
\end{itemize}
By symmetry, the next state for $v_3$ is \textsc{Gnash} and for $v_4$ is \textsc{Glory}. The resulting state is $s_1$. An identical calculation starting from $s_1$ shows the next state is $s_0$, confirming the two-cycle.
\end{proof}

\subsection{\texttt{HeavenExample}}
This model adds a hub $g=h$ to the 4-cycle. The hub has outgoing edges of weight $W$ to each of $\{v_1, \dots, v_4\}$.

\begin{lemma}[Hub and Rest Weights]
For any non-hub vertex $v_i \in \{v_1, v_2, v_3, v_4\}$:
\begin{enumerate}[nolistsep, label=(\roman*)]
    \item $\text{hub\_weight}(v_i) = W$.
    \item $\text{rest\_weight}(v_i) = 2$.
\end{enumerate}
\end{lemma}
\begin{proof}\hfill
\begin{enumerate}[nolistsep, label=(\roman*)]
    \item By definition, $\text{hub\_weight}(v_i) = w(h, v_i) = W$.
    \item By definition, $\text{rest\_weight}(v_i) = \sum_{u \neq h} w(u, v_i)$. The non-hub neighbours of $v_i$ are its two neighbours on the cycle, each contributing weight 1. The sum is $1+1=2$.
\end{enumerate}
\end{proof}

\begin{theorem}[Counterexample for Weak Hub]
\label{thm:heaven_counterexample}
If $W < 2$, then for any non-hub vertex $v_i$, $\text{next}_{\text{heaven}}(\text{all\_gnash}, v_i) = \textsc{Gnash}$.
\end{theorem}
\begin{proof}
By \autoref{lem:domination_all_gnash}, the next state is \textsc{Glory} iff $\text{hub\_weight}(v_i) \ge \text{rest\_weight}(v_i)$. From the preceding lemma, this is $W \ge 2$. Since $W < 2$ by hypothesis, this condition is false, so the next state must be \textsc{Gnash}.
\end{proof}

\begin{theorem}[Convergence for Strong Hub]
\label{thm:heaven_convergence}
If $W \ge 2$, then for any state $s$ and any vertex $v$, $\text{next}_{\text{heaven}}(s, v) = \textsc{Glory}$.
\end{theorem}
\begin{proof}
We apply \autoref{thm:global_conv_equiv}, which requires verifying that $\forall v \neq g, \text{hub\_weight}(v) \ge \text{rest\_weight}(v)$. For any non-hub vertex $v_i$, this condition is $W \ge 2$, which holds by hypothesis. Therefore, global one-step convergence is guaranteed.
\end{proof}

\end{document}